\documentclass{PoS}

\usepackage{graphicx,subfigure}

\title{A QCD analysis for nuclear PDFs at NNLO}

\ShortTitle{A QCD analysis for nuclear PDFs at NNLO}

\author{\speaker{Marina Walt}\\
        Institute for Theoretical Physics, T\"ubingen University, Auf der Morgenstelle 14, 72076 T\"ubingen, Germany \\
        E-mail: \email{marina.walt@uni-tuebingen.de}}

\author{Ilkka Helenius\\
        University of Jyvaskyla, Department of Physics, P.O. Box 35, FI-40014 University of Jyvaskyla, Finland \\
        Helsinki Institute of Physics, P.O. Box 64, FI-00014 University of Helsinki, Finland \\
        E-mail: \email{ilkka.m.helenius@jyu.fi}}
        
\author{Werner Vogelsang\\
        Institute for Theoretical Physics, T\"ubingen University, Auf der Morgenstelle 14, 72076 T\"ubingen, Germany \\
        E-mail: \email{werner.vogelsang@uni-tuebingen.de}}        
        
\abstract{A new QCD analysis for nuclear parton distribution functions (nPDFs) at next-to-leading order (NLO) and next-to-next-to-leading order (NNLO) is presented. The framework of the analysis, including the form of the parameterization as well as the included DIS data sets, are discussed. The results of this QCD analysis are compared to the existing nPDF sets and to the fitted data. The presented framework is based on an open-source tool, \textsc{xFitter}, which has been modified to be applicable also for a nuclear PDF analysis. The required modifications are covered as well. Finally, an outlook for the next developments of the QCD analysis for nuclear PDFs is given.}

\FullConference{XXVII International Workshop on Deep-Inelastic Scattering and Related Subjects - DIS2019\\
		8-12 April, 2019\\
		Torino, Italy}

\begin{document}

\section{Introduction}

The purpose of nuclear parton distribution functions (nPDFs) is to describe the collinear momentum distribution of the partons (quarks and gluons) inside a proton which is bound to a nucleus. The knowledge of nuclear parton distribution functions is relevant for heavy-ion experiments at the LHC and at RHIC to analyse and interpret the measurements. The alignment of theoretical calculations, i.e. computed cross sections, to the experimentally obtained data is key for making predictions for the future projects, like for example the electron-ion collider (EIC) \cite{eic}. The fundamental interactions between the partons are described with quantum-chromodynamics (QCD). According to the collinear factorization theorem \cite{factorization}, the perturbatively calculable partonic scattering processes can be factorized from the non-perturbative PDFs. The PDFs cannot be calculated from the first principles QCD but their scale evolution can be derived from perturbative QCD. Therefore the nPDFs can be derived in a QCD analysis by applying suitable data for bound nucleons.

\section{Theoretical basis} 

The analysis has been performed at next-to-leading order (NLO) and next-to-next-to-leading order (NNLO) in perturbative QCD. The order of perturbative theory affects two parts of the analysis procedure. First, it defines at which order the DGLAP evolution in $Q^2$ is performed, i.e. the powers of $\alpha_{\mathrm{S}}$ taken into account in splitting functions (e.g. $\alpha_{\mathrm{S}}^3$ for NNLO \cite{splittingf-nnlo1,splittingf-nnlo2}). Second, it defines the precision at which the partonic cross sections are computed. In this analysis we use data for neutral and charged current deeply inelastic scattering (DIS), where the appropriate QCD corrections are effectively included in the definitions of the structure functions $F_i$, see e.g. Ref.~\cite{f2nnlo} for $F_2$ at NNLO.

The nuclear PDFs are often determined based on a specific, existing \textit{free} proton PDF set. In this analysis, we first determine our own free proton baseline using DIS data from HERA, BCDMS and NMC experiments. For the basic form of the PDF parameterization at the initial scale of the analysis the ansatz 
\begin{equation}
xf^{p/A}_i\left(x,Q_0^2 \right) = c_0\,x^{c_1} (1-x)^{c_2} \left(1+c_3\,x + c_4\,x^2 \right)
\label{eq:pdf-parameterization}
\end{equation}
with $i=g,\,u_v\,d_v\,\bar{u},\,\bar{d},\,s,\,\bar{s}$ is used. A similar ansatz has been used to derive the \mbox{HERAPDF2.0}~\cite{hera20} proton set. The same form of the parameterization (\ref{eq:pdf-parameterization}) is valid for both, proton and nuclear PDFs. The difference appears in regards to the parameters $c_i$ ($i=0,...,4$). For nuclear PDFs the coefficients in equation (\ref{eq:pdf-parameterization}) are further parameterized to be dependent on the nuclear mass number $A$ as
\begin{equation}
c_k\,\rightarrow c_k(A) = c_{k,0}+c_{k,1}\left( 1 - A^{-c_{k,2}} \right)
\label{eq:coeff-A}
\end{equation}
with $k={0,\dots,4}$. This form of $A$-dependent coefficients was used in the nCTEQ15 analysis \cite{nCTEQ15}. This $A$-dependent parameterization has the advantage that in case of a free proton $(A=1)$ the term $\left( 1 - A^{-c_{k,2}} \right)$ in equation (\ref{eq:coeff-A}) becomes zero and the functional form of a free proton is automatically retained.

The nuclear parton distribution function $f_i^{\,N/A}$ for a \textit{bound} nucleon inside a nucleus with mass number $A$ is constructed from the \textit{bound} proton's PDF $f_i^{\,p/A}$ (not from a free proton's PDF $f^p$). In particular for the distribution of partons in a bound nucleon we write
\begin{equation}
f_i^{\,N/A} \left( x,Q^{\,2} \right) = \frac{Z\cdot f_i^{\,p/A}+ (A-Z)\cdot f_i^{\,n/A}}{A}\,,
\label{eq:nucleon}
\end{equation} 
where the $Z$ is the number of protons in the nucleus. The PDF of the \textit{bound} neutron $f_i^{\,n/A}$ is determined from the fitted proton's PDF using the isospin symmetry. 

As can be seen from equation (\ref{eq:nucleon}), if $Z\neq \frac{A}{2}$, the fraction of proton's PDF $f_i^{\,p/A}$ and the one of neutron's PDF $f_i^{\,n/A}$ become different for different combinations of $A$ and $Z$. However, sometimes the experimental collaborations apply so-called isoscalar corrections on the measure data, so that $Z=(A-Z)=\frac{A}{2}$ can be used for the PDF decomposition of a nucleus. As there is no need to use such a decomposition in the analysis such a simplification is not required here, but the isoscalar corrections need to be reverted in order to be consistent with the given measurement. 

For the nuclear part of this QCD analysis the coefficients $c_{k,0}$ (equation (\ref{eq:coeff-A})) for all flavors were kept fixed based on the precedent proton PDF analysis. As part of the nuclear PDFs only the so-called nuclear parameters $c_{k,1}$ and $c_{k,2}$ were fitted for different flavors. For the flavor decomposition $u_v\neq d_v$ has been allowed for the valence quarks, and $\bar{u}=\bar{d}=s=\bar{s}$ is assumed for the sea quarks. Furthermore, the number sum rule and the momentum sum rule are used to constrain the normalizations of $d_v$, $u_v$ and $\bar{u}$. In total, $16$ free nuclear parameters have been fitted as part of this QCD analysis.

\section{Analysis framework}

The fitting framework is based on an open-source tool \textsc{xFitter} \cite{xfitter-project,xfitter200} which has been modified to be applicable also for a nuclear PDF analysis. First, a new PDF type 'nucleus' has been introduced. If the mass number $A$ and the proton number $Z$ are set to $A=1$ and $Z=1$, the new PDF type 'nucleus' and the existing PDF type 'proton' coincide. Next, an explicit $A$-dependence (cf. eq. \ref{eq:coeff-A}) has been implemented for the fitted coefficients. In order to build a nucleus or a \textit{bound} nucleon (cf. eq. \ref{eq:nucleon}) the parton flavor decomposition has been modified accordingly. For that, the isospin symmetry is assumed. 

A set of necessary modifications results from the fact that the measured quantities are provided in form of ratios, instead of absolute cross sections. For example, often the experimental data is published for a ratio of a cross section measured on one nucleus with mass number $A_1$ to the cross section of the other nuclear target $A_2$, i.e. $\sigma(A_1)/\sigma(A_2)$ for cross sections or $F_2(A_1)/F_2(A_2)$ for structure functions. Thus, the analysis routine was modified to reflect the information if the theoretical predictions need to be compared to an absolute quantity or to a ratio (CInfo$=$'ratio').

Besides that, some experiments apply isoscalar corrections to the measured data and publish only the modified information. Thus, the analysis procedure needs to be adapted so that the calculated quantities are consistent with the iso-corrected experimental data. For this purpose, different flags were introduced in \textsc{xFitter} for the different forms of isoscalar corrections, which are specific to the corresponding experiments (CInfo$=$'NMC', 'EMC', 'SLAC').

Eventually, another modification on \textsc{xFitter} was necessary for the treatment of charged current DIS processes measured in neutrino-nucleus scattering reactions. As part of this framework, the differential cross sections $\mathrm{d}\sigma^2/\mathrm{d}y\mathrm{d}Q$ (instead of the structure functions $F_2$, $F_3$ as in Ref.~\cite{dssz}) were used for the analysis. Thus, new reactions 'neutrino+p CC' and 'antineutrino+p CC' have been implemented in \textsc{xFitter}.

\section{Results}

\begin{figure}[tb!]
\begin{center}
      \subfigure{        
              \includegraphics[width=0.32\textwidth]{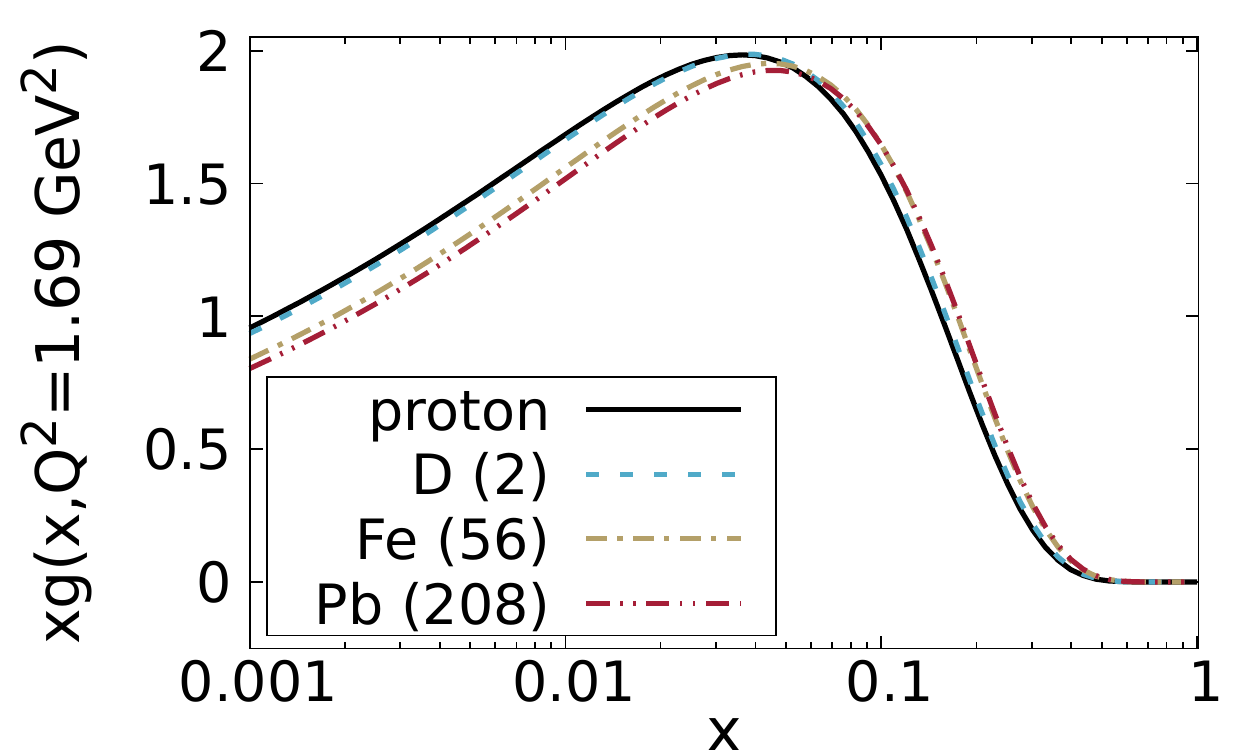}} 
      \subfigure{        
              \includegraphics[width=0.32\textwidth]{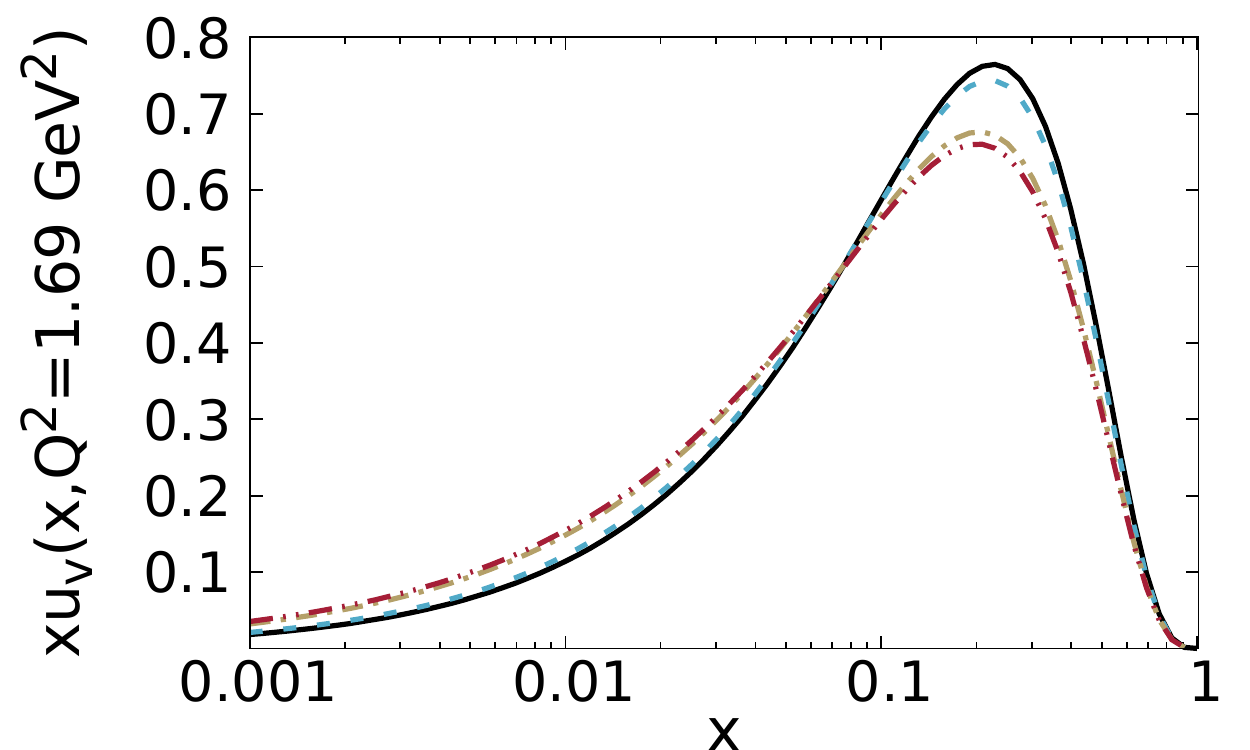}} 
       \subfigure{        
              \includegraphics[width=0.32\textwidth]{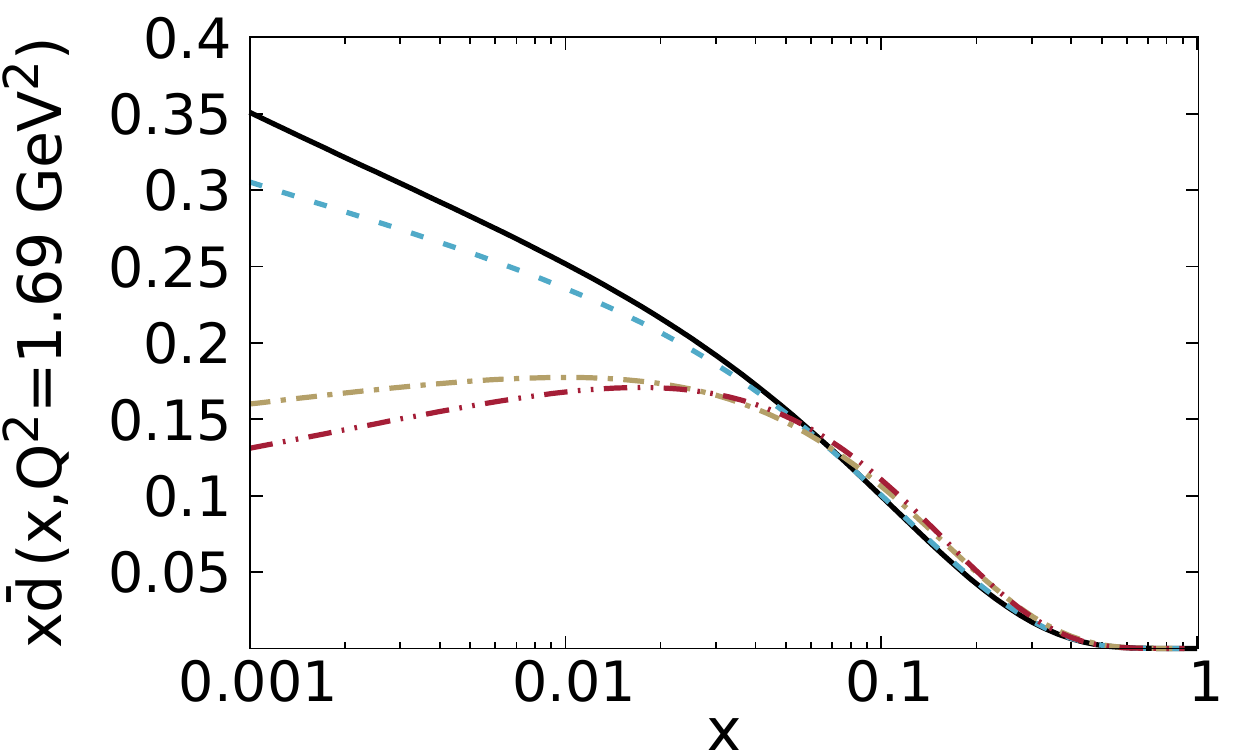}} 
\end{center} 
\caption{Preliminary nPDF results at next-to-next-to-leading order (NNLO) for different nuclei at the initial scale $Q_0^2=1.69~\mathrm{GeV^2}$. The solid black line shows the distribution function of a \textit{free} proton. The dotted colored lines represent the distribution functions of protons bound in different nuclei (here: deuteron 'D', iron 'Fe' and lead 'Pb'). The corresponding mass number $A$ is provided in brakets.}
\label{fig-pdfs}
\end{figure}

The preliminary nPDF results at NNLO for different nuclei at the initial scale $Q_0^2=1.69~\mathrm{GeV^2}$  are shown in figure \ref{fig-pdfs}. The difference of gluon distributions for different nuclei is found small at NNLO. The valence quark distributions (here $u_v$) vary a bit more for the different nuclei. The last subfigure on the right-hand side shows that the variance in the amplitude for the sea quark distributions (here $\bar{d}$ but equal for all flavors) is quite large at NNLO. Additionally, the results in figure \ref{fig-pdfs} show that the major contribution by valence quarks is in the large $x$ region, whereas the occupation by sea quarks is higher at the small $x$ scale, as expected.

\begin{figure}[tb!]
\begin{center}
      \subfigure{        
              \includegraphics[width=0.32\textwidth]{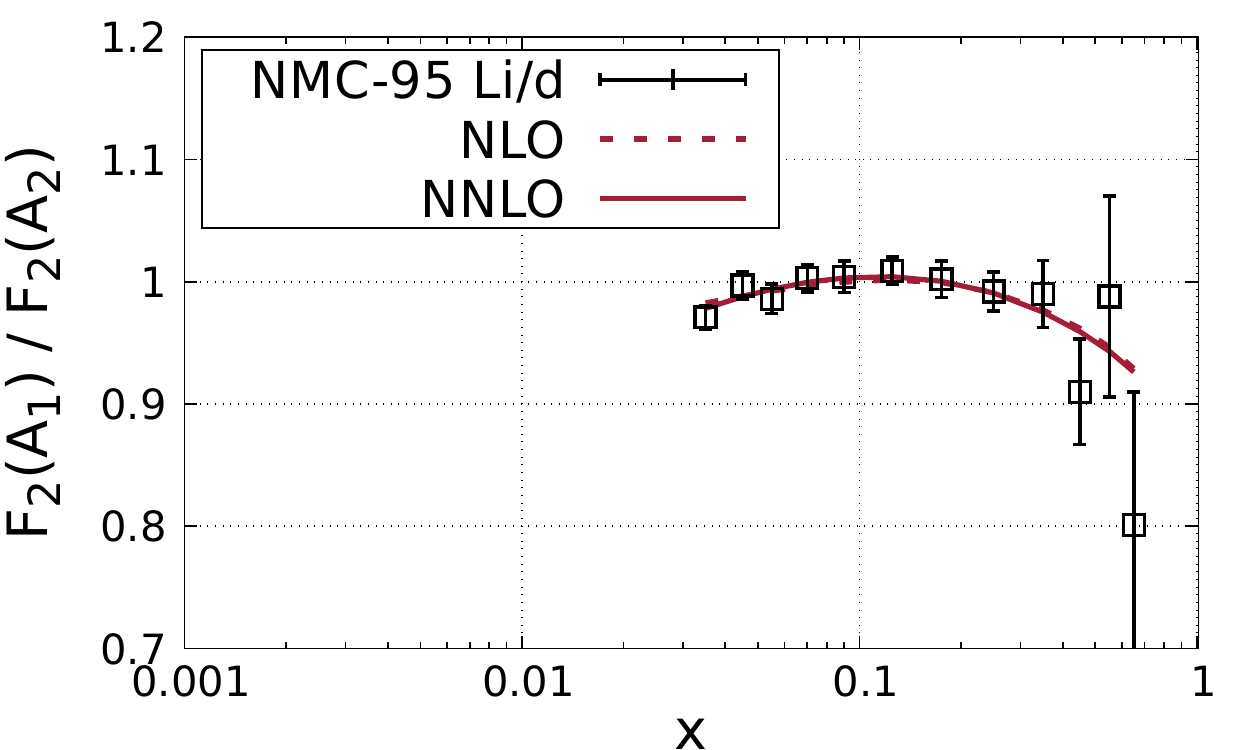}} 
      \subfigure{        
              \includegraphics[width=0.32\textwidth]{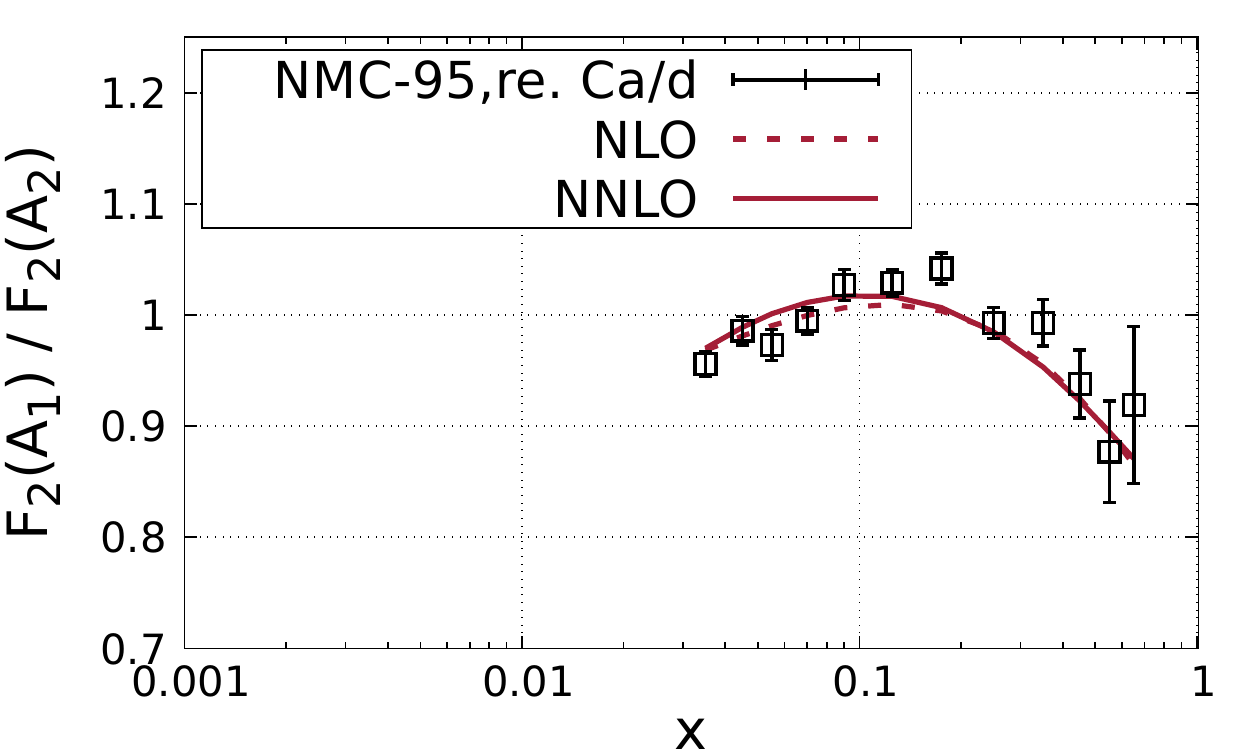}} 
       \subfigure{        
              \includegraphics[width=0.32\textwidth]{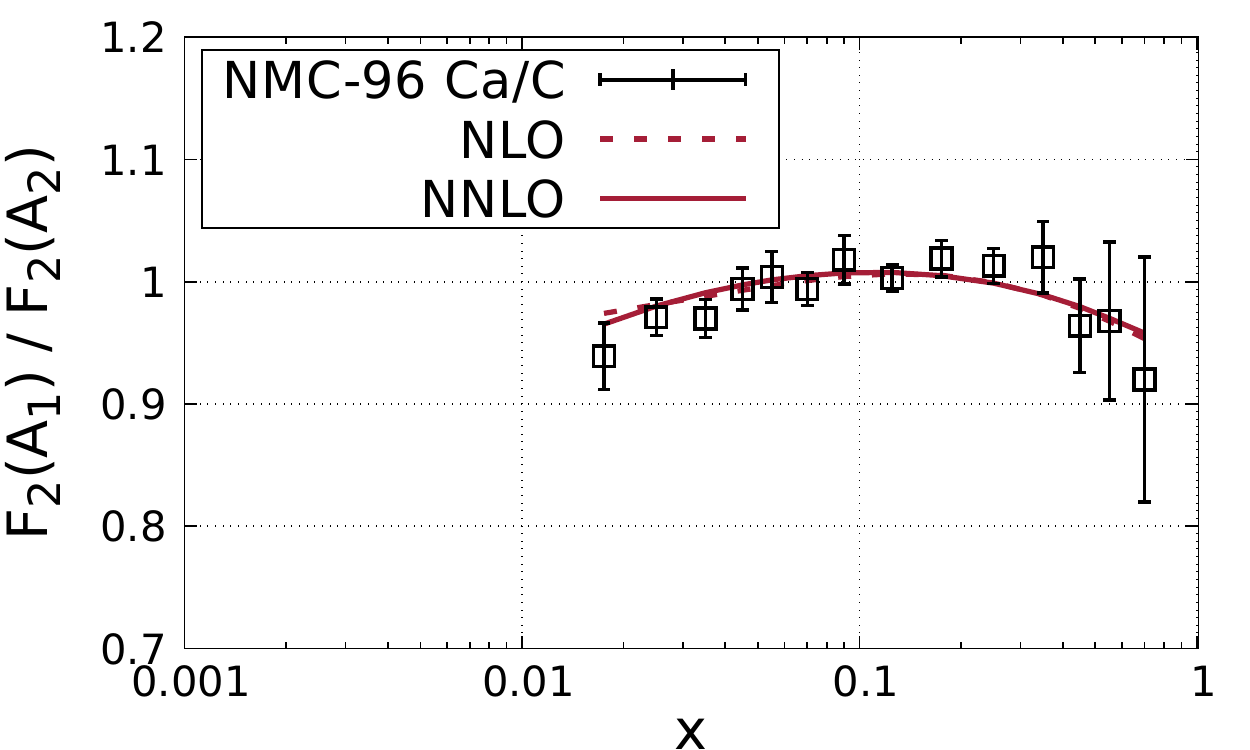}}                                          
      \subfigure{        
              \includegraphics[width=0.32\textwidth]{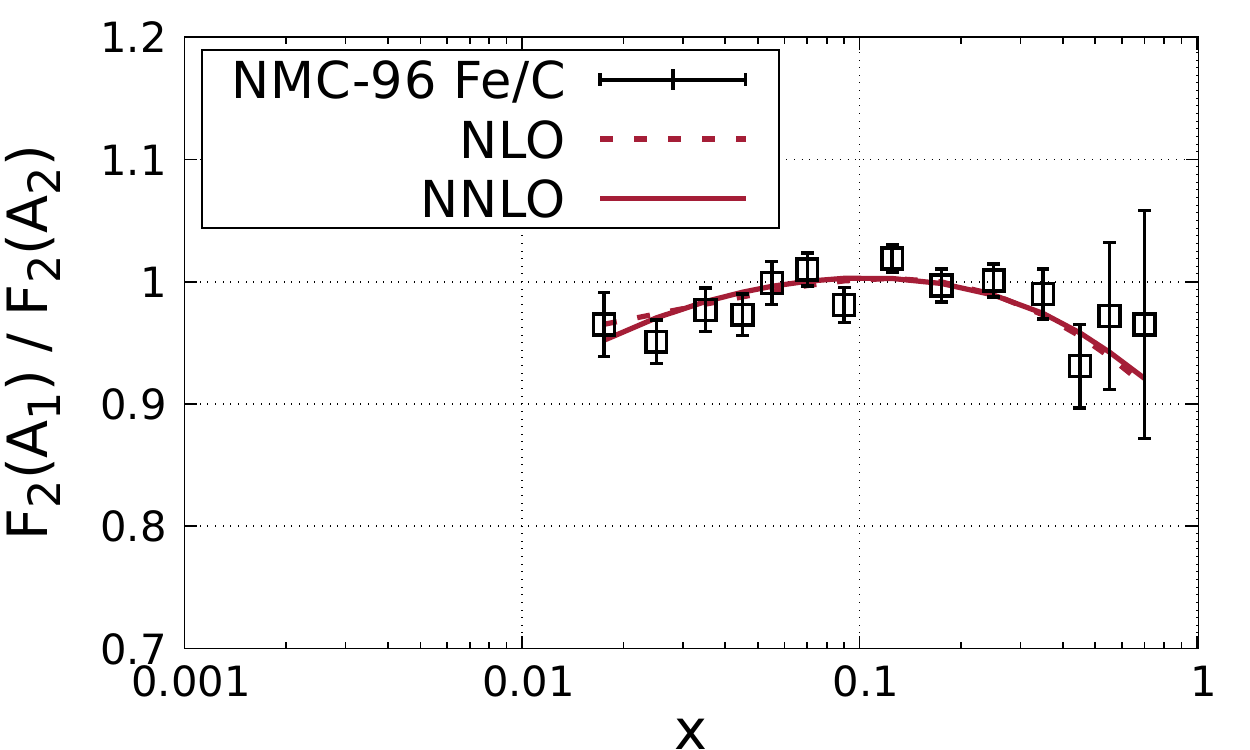}} 
      \subfigure{        
              \includegraphics[width=0.32\textwidth]{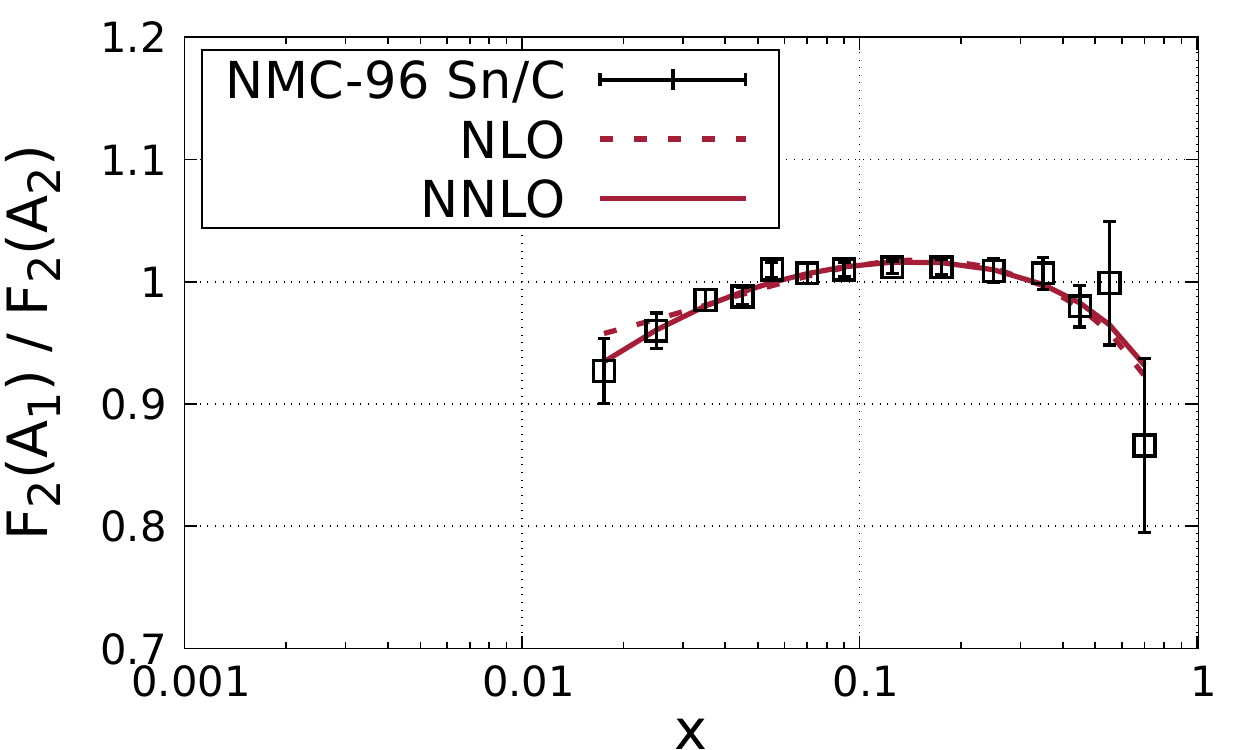}} 
      \subfigure{        
              \includegraphics[width=0.32\textwidth]{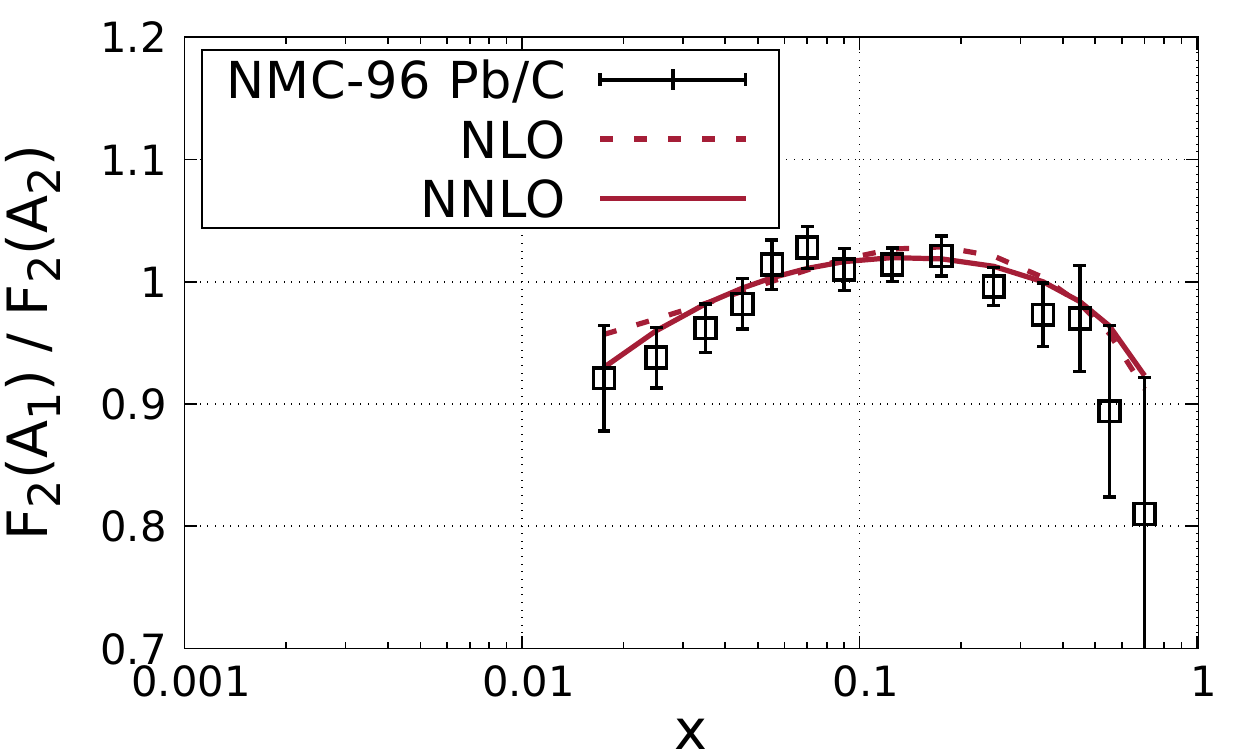}}                                           
\end{center} 
\caption{Comparison of the NLO (solid) and NNLO (dashed) analysis to the experimental data for a selected representative subset of the applied data for the neutral current DIS process.}
\label{fig-dis}
\end{figure}

\begin{figure}[tb!]
\begin{center}
      \subfigure{        
              \includegraphics[width=0.32\textwidth]{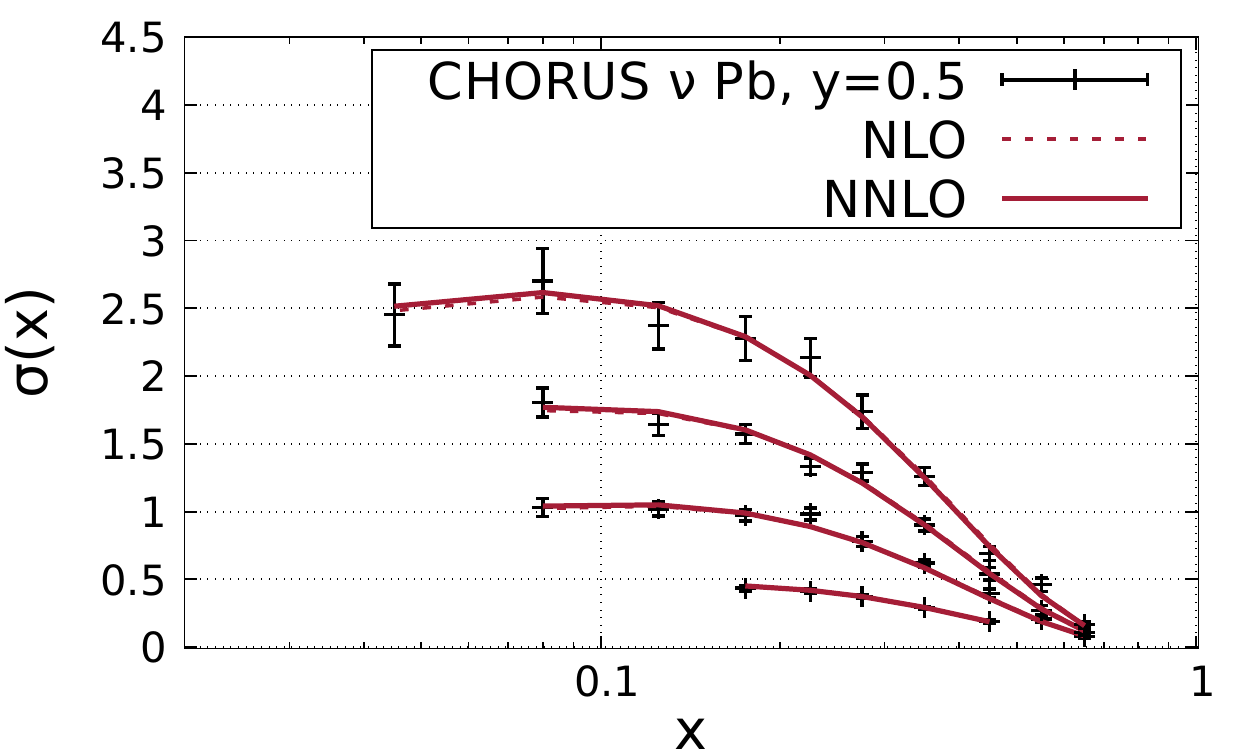}} 
      \subfigure{        
              \includegraphics[width=0.32\textwidth]{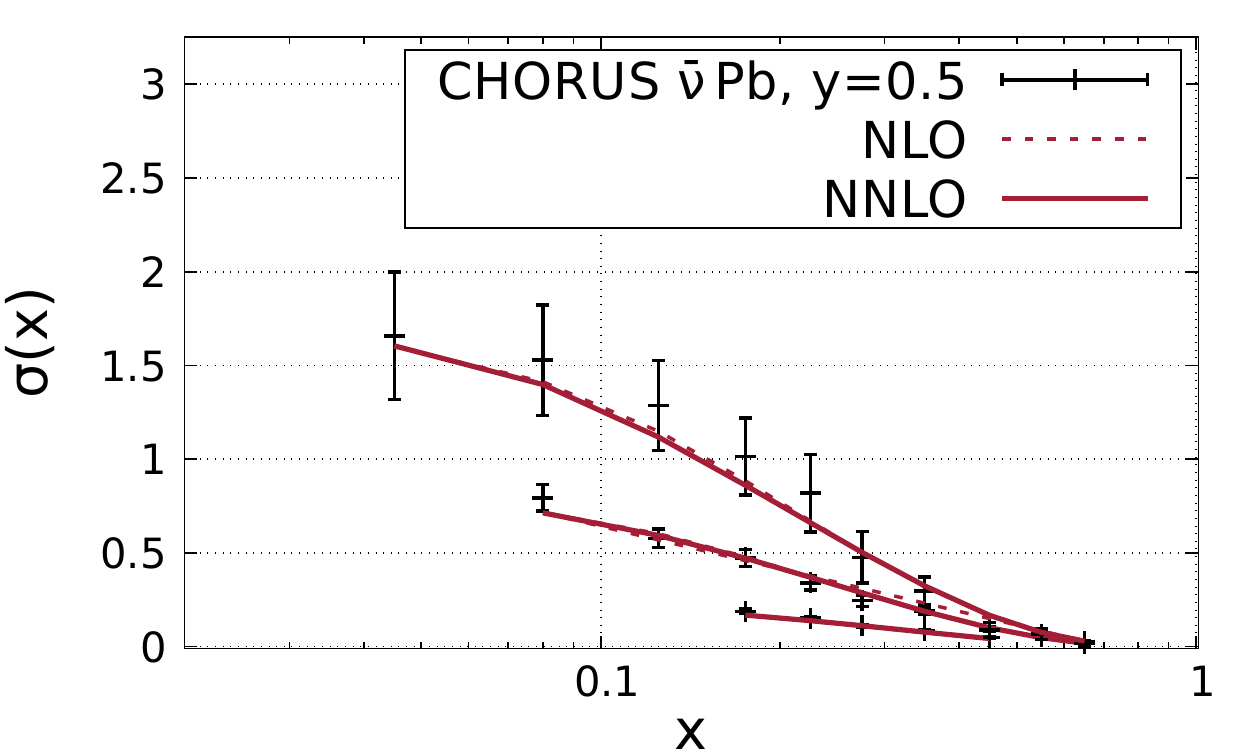}} 
       \subfigure{        
              \includegraphics[width=0.32\textwidth]{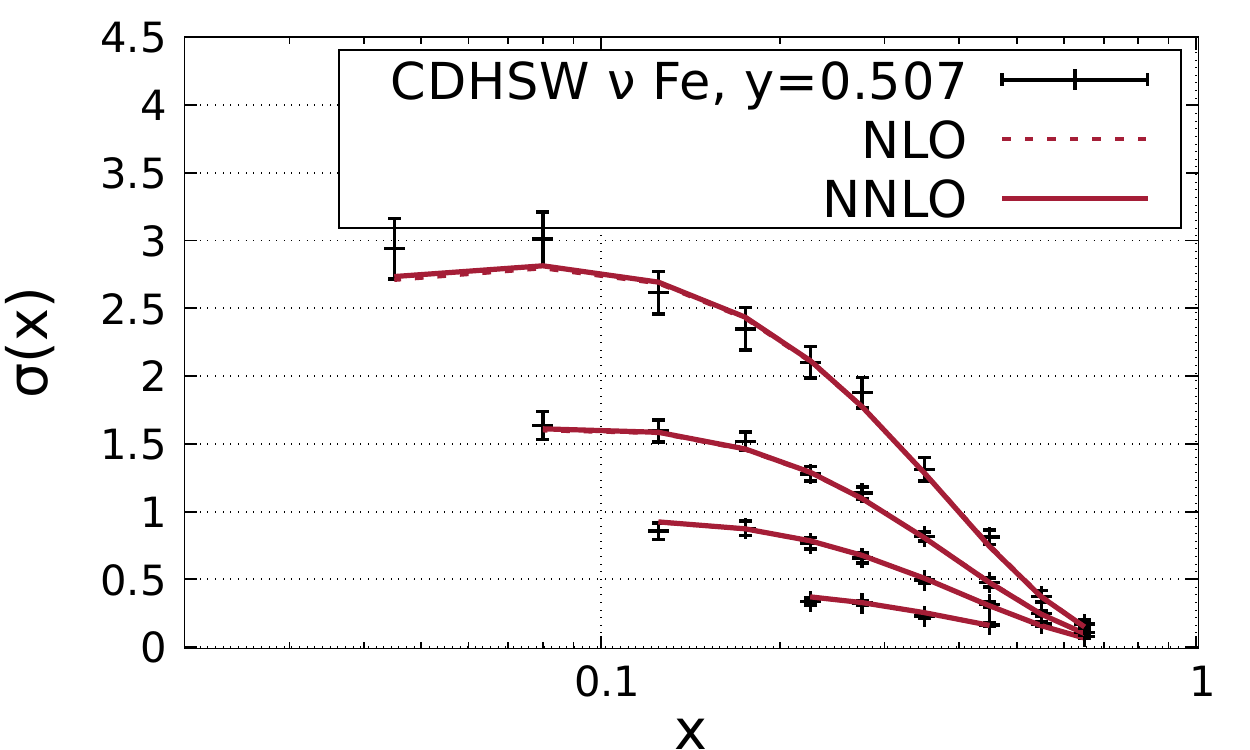}}                                          

\end{center} 
\caption{Comparison of the NLO (solid) and NNLO (dashed) analysis to the experimental data measured in neutrino-nucleus scattering for the charged-current DIS process with $y=0.5$.}
\label{fig-neutrinos}
\end{figure}

A comparison of the obtained cross sections to the experimental data is shown in figure \ref{fig-dis} and figure \ref{fig-neutrinos} for a selected representative subset of the applied data. For the complete information please refer to our forthcoming publication \cite{tuju19}. As can be seen in figure \ref{fig-dis} and figure \ref{fig-neutrinos}, the agreement of the calculated quantities with the measurements is very good at NLO and NNLO. This implies that the qualities of the QCD analyses at NLO and NNLO are comparable for the available constraints and within the given experimental uncertainties. Furthermore, figure \ref{fig-neutrinos} shows, that experimental data from neutral-current DIS processes and charged-current neutrino-nucleus DIS processes were included successfully in a common fit. 

\begin{figure}[tb!]
\begin{center}
\subfigure{\includegraphics[width=0.32\textwidth]{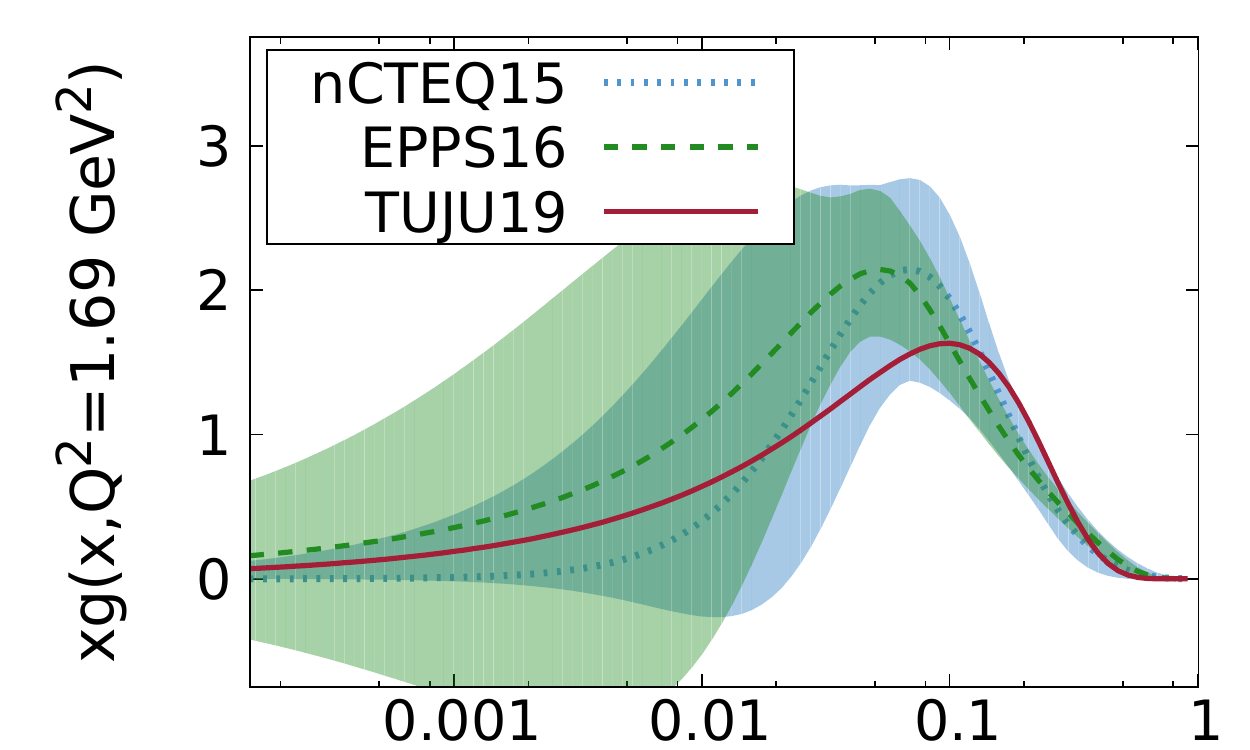}} 
\subfigure{\includegraphics[width=0.32\textwidth]{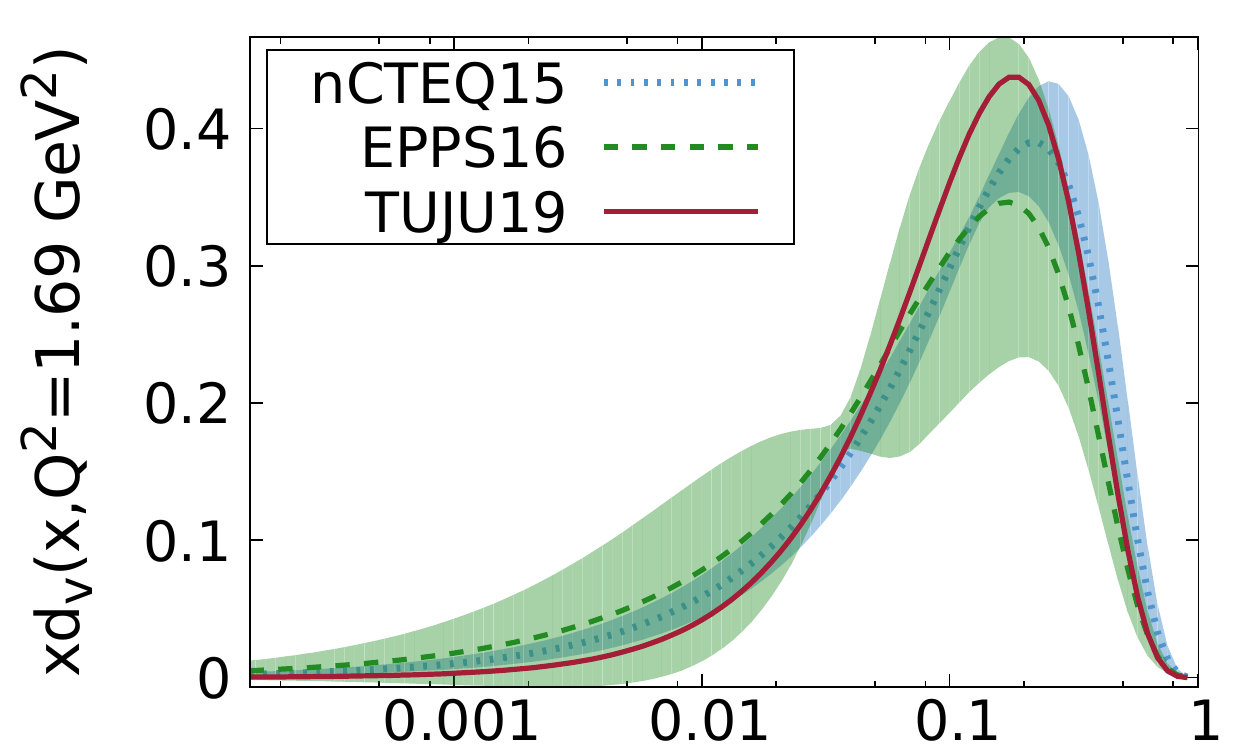}}
\subfigure{\includegraphics[width=0.32\textwidth]{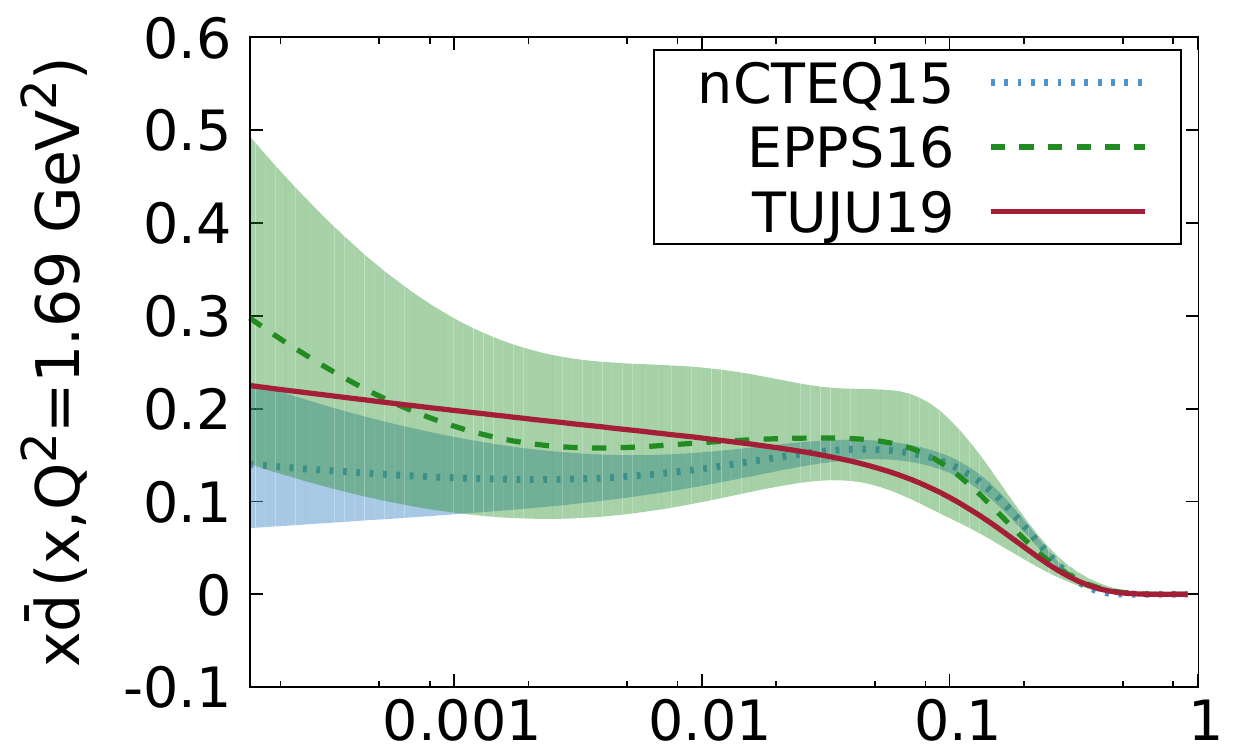}}
\subfigure{\includegraphics[width=0.32\textwidth]{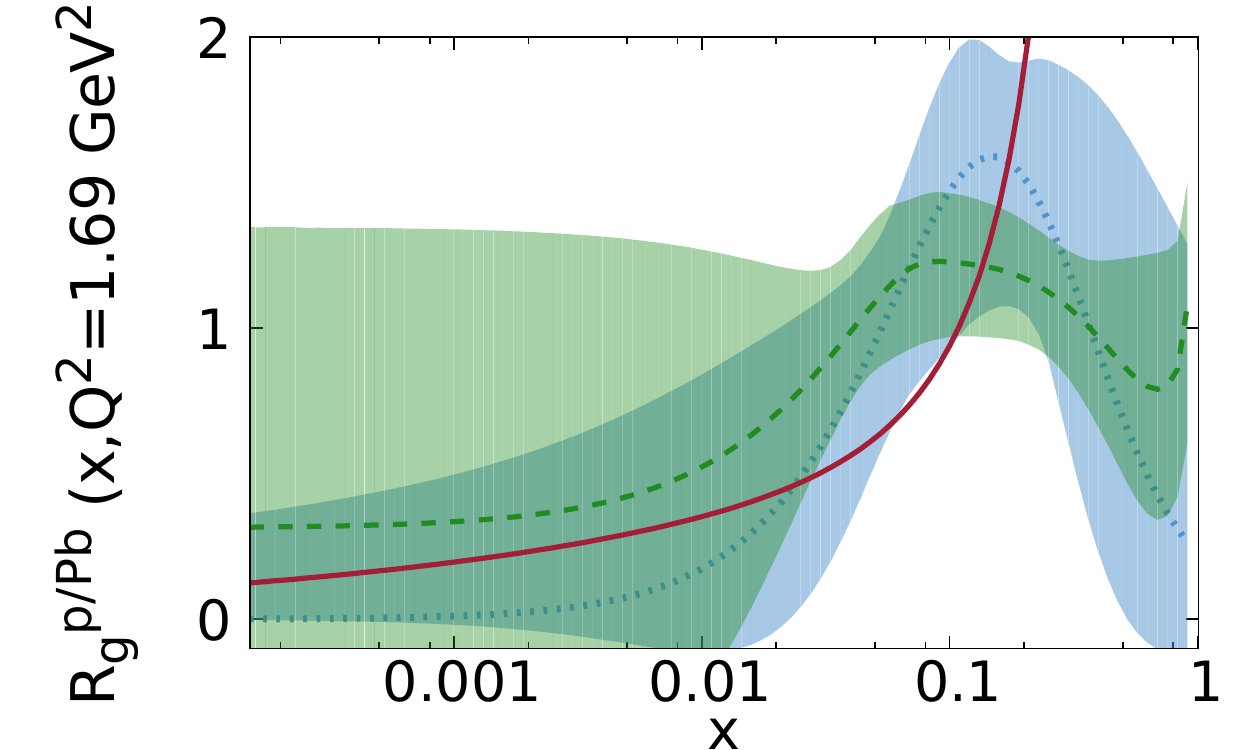}}
\subfigure{\includegraphics[width=0.32\textwidth]{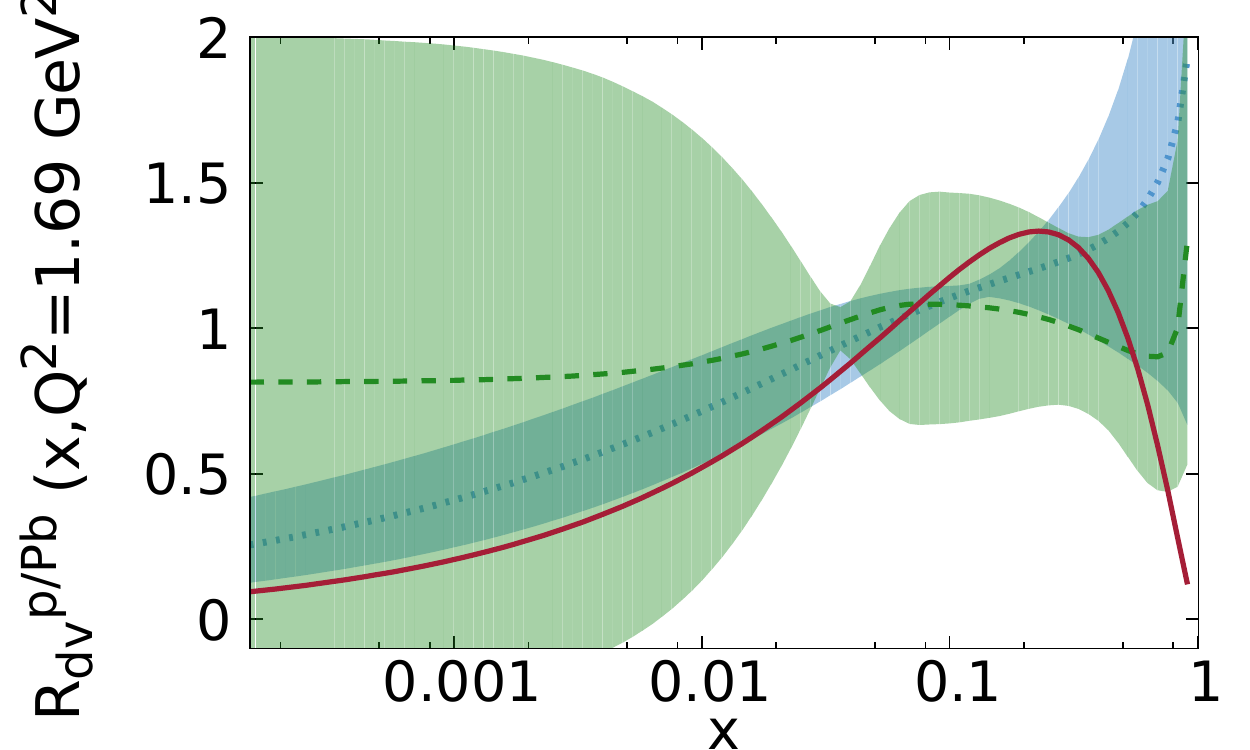}}
\subfigure{\includegraphics[width=0.32\textwidth]{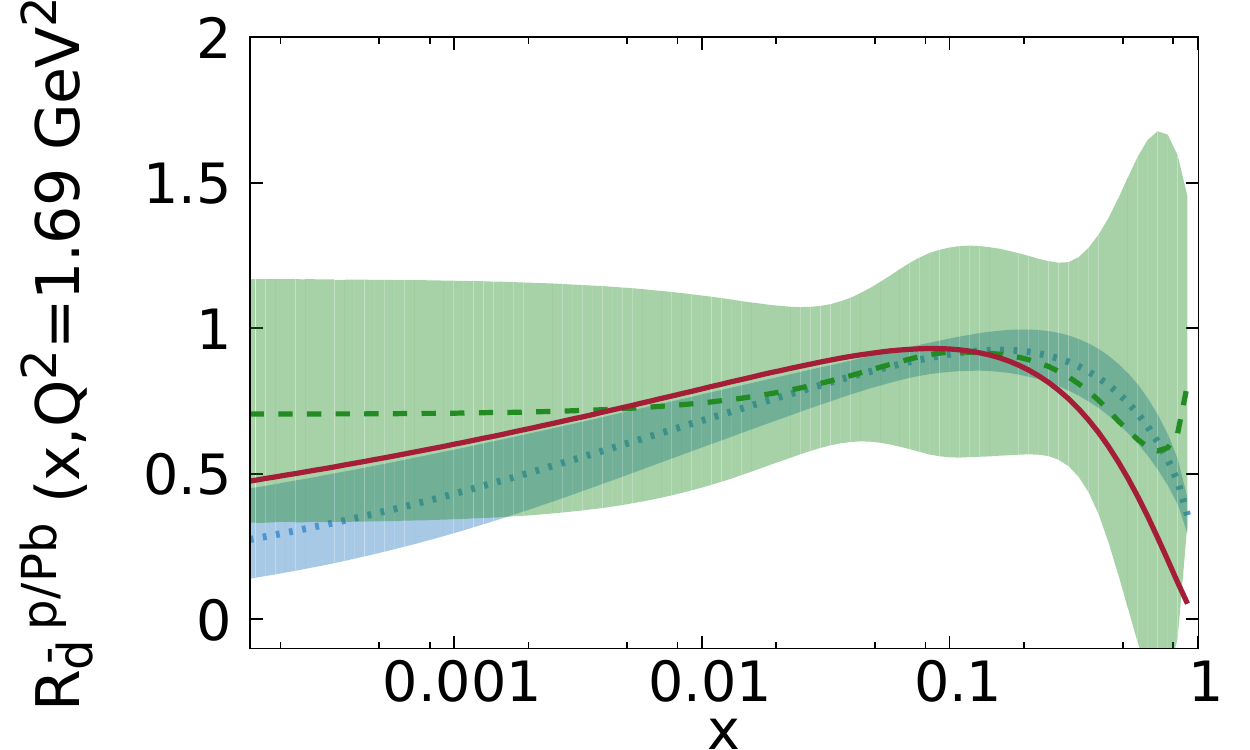}}
\end{center}
\caption{Comparison of central parton distribution functions of this framework $\left(\mathrm{TUJU19}\right)$ to the available LHAPDF sets nCTEQ15 and EPPS16 at NLO for a bound proton in lead (Pb). In the upper line, parton distribution functions are shown. In the second row a ratio of a proton in lead compared to a free proton is presented.}
\label{comparison-npdfs}
\end{figure}
The central values of the nPDFs obtained as part of this work $\left(\mathrm{TUJU19}\right)$ are compared to nCTEQ15 \cite{nCTEQ15} and EPPS16 \cite{epps16} fits at NLO in figure \ref{comparison-npdfs}. Besides, other recent nPDF analyses have been performed by different collaborations, including DSSZ \cite{dssz} at NLO, and KA15 \cite{ka15} and nNNPDF1.0~\cite{nnnpdf} at NNLO. In addition to the absolute parton distribution functions shown in figure \ref{comparison-npdfs}, also the ratios $R_i^{\,p/Pb}~=~xf_i^{\,p/Pb}(x,Q^2)~/~xf_i^{\,p}(x,Q^2)$ of a proton in lead compared to a free proton per parton flavor $i=\mathrm{g},\,\mathrm{d_v}\,,\mathrm{\bar{d}}$ are presented. As can be seen, the central PDFs are mostly within the error bands of the other sets. Only the gluon nuclear modification at large-$x$ deviates from the previous analyses at the initial scale $Q_0^2$, but agreement is found at higher scales. The error bands and further details on the uncertainty analysis will be presented in our forthcoming publication \cite{tuju19}.

\section{Summary and outlook}

A new QCD analysis for nuclear parton distribution functions at NLO and NNLO is presented, referred to as TUJU19. In the first phase, experimental data from the measurements of neutral current DIS processes and charged current neutrino-nucleus DIS have been included. The obtained results of this QCD analysis show a nice agreement with the existing nPDF sets and the fitted data. Rather than choosing an already existing set of proton PDFs as a baseline for the nuclear PDFs, we have developed our own proton set. Furthermore, deuteron has been considered being a nucleus with non-negligible nuclear effects. The numerical setup is based on the open-source tool \textsc{xFitter} which has been modified to be applicable for nuclear PDF analyses. In the next phase we plan to include experimental data for Drell-Yan processes. As a Long-term goal, an inclusion of further data from RHIC and LHC experiments, e.g. for jets and W, Z bosons, is foreseen. 

\section{Acknowledgments}
This work was supported in part by the Bundesministerium f\"{u}r Bildung und Forschung (BMBF) grant 05P18VTCA1. The authors acknowledge support by the state of Baden-W\"{u}rttemberg through bwHPC.

\end{document}